\documentclass[preprint,aps,amsmath,eqsecnum,fleqn]{revtex4}
\usepackage{bm}
\usepackage[dvips]{graphicx}
\usepackage{color}
\usepackage{latexsym}
\usepackage{eufrak}
\usepackage{amsmath}

\newcommand{\eps}{\epsilon}

\begin{document}

\title{In-plane spectrum in superlattices}

\author{B. Laikhtman}
 \email{borisl@atrg.com}
\affiliation{Racah Institute of Physics, Hebrew University, Jerusalem 91904 Israel}
\affiliation{State University of New York at Stony Brook, Stony Brook, NY 11794, USA}

\author{S. Suchalkin and G. Belenky}
\affiliation{State University of New York at Stony Brook, Stony Brook, NY 11794, USA}

\date{\today}

\begin{abstract}
We show that the existing theory does not give correct in-plane spectrum of superlattices at small in-plane momentum. Magneto-absorption experiments demonstrate that the energy range of the parabolic region of the spectrum near the electron subband bottom is by the order of magnitude lower than the value predicted by the traditional approach. We developed a modified theory according to which the energy range of the parabolic region and carrier in-plane effective masses are determined by the effective bandgap of the superlattice rather than by the bulk bandgaps of the superlattice layers. The results of the new theory are consistent with the experiment.

\end{abstract}

\pacs{Valid PACS appear here}
\maketitle

\section{Introduction}

Methods used for calculation of carrier spectrum in heterostructures can be roughly divided in two big classes: analytical and numerical. An advantage of analytical methods is that they present the whole spectrum in a comprehensive form that makes it easy to analyse its features and effect of different factors. The purpose of the present paper is to show that an analytical method typically used for calculation of superlattice (SL) carrier spectrum does not work for in-plane spectrum. We modify the method to obtain correct results.

The work was motivated by recent experiments with InAsSb$_{x}$/InAsSb$_{y}$ short period SLs. Extremely narrow band gap found in these SLs\cite{Belenky15} makes them important for fabrication of far infrared detectors and light emitting diodes that have potential applications such as pollutant gas sensing, molecular spectroscopy, process monitoring, disease analysis and infrared scene projection.\cite{Crowder02,Zotova06,Koerperick11}
The cyclotron resonance measurements in these SLs found out linear in-plane electron spectrum starting from unusually small energy of about 10 meV.\cite{Suchalkin}

In general, nonparabolicity of electron spectrum is well known. It was studied theoretically in both quantum wells and SLs\cite{Bastard81,Welch84,Hiroshima86,Nelson87,Yoo89,Winkler96,Wetzel96} and was detected in cyclotron resonance measurements in quantum wells.\cite{Yang93,Scriba93} This nonparabolicity becomes substantial at energies around band gap of the
material of the quantum well which is a few hundered meV and it was explained in the frame of the Kane model.\cite{Scriba93,Yang93,Winkler96,Wetzel96} Theory also predicts nonparabolicity at small energy energy scale in
superlattices fabricated of narrow gap materials like HgTe-CdTe.\cite{Gerchikov92}

A special feature of the new experiment is that the band gap in constituent layers of InAsSb is more than 100 meV.\cite{Adachi,Vurgaftman01} According to regular theory one can expect that the in-plane spectrum becomes non-parabolic at this energy scale.  It appears, however, that the nonparabolicity becomes large at energies by the order of magnitude smaller. This discrepancy requires a thorough analysis of the theory foundation.

Even a short view at the existing theory brings about some doubts about its applicability to the in-plane carrier spectrum in SLs. This theory is based on effective mass approximation in each layer with effective masses equal to their values in the respective bulk materials. So it is assumed that $\bm{kp}$ method used as a foundation for effective mass approximation is equally applied in bulk and SL layers. Meanwhile, if the in-plane vector in SL, $\bm{k}_{\parallel}$, is very small the part of the $\bm{kp}$ Hamiltonian ${\cal H}_{\parallel}$ containing $\bm{k}_{\parallel}$ can be considered as a perturbation. In this case the calculation of the SL spectrum is broken in two steps. At the first step effective masses of separate layers and SL spectrum in the growth direction with $\bm{k}_{\parallel}=0$ are obtained in regular way. The second step is calculation of the in-plane spectrum with perturbation theory. This means that the in-plane spectrum is formed on the basis of SL spectrum in the growth direction but not spectra of separate layers. This difference is very important because the value of effective mass is crucially dependent on the gap between the conduction and valence band. In SL those bands are formed by two different materials and their positions depend on material band offsets and parameters of the structure. As a result, the gap may appear much smaller than in each material separately. This leads to decrease of the effective mass and shrinking of the parabolic region of the in-plane spectrum.

In the present paper we develop a new approach to spectrum calculation and show the difference of the results between the old and new methods. A special feature of our approach compared to previous works is that we consider ${\cal H}_{\parallel}$ as a perturbation, that is justified in many cases. That is the in-plane spectrum is substantially modified due to penetration of wave functions from wells to barriers. Although such penetration was studied in earlier works, it was considered as a mixture of two spectra with different effective masses.\cite{Yang93,Wetzel96} We show that it is necessary fist to mix bands of different materials at $\bm{k}_{\parallel}=0$ and then calculate the in-plane spectrum.

To make our arguments and calculation more simple and clear we neglect some details that for our purpose have secondary importance. So in $\bm{kp}$ Hamiltonian we include only coupling between conduction and valence bands and neglect coupling with remote bands. We discard also spin-orbit split band. The result of such calculation gives reliable estimates of spectrum characteristics and their dependence on relevant parameters but may be not very precise in their numerical values and miss some subtle details as, e.g., warping of the hole spectrum.

Our approach is basically analytical. We use effective mass approximation when it is justified and obtain analytic results when this is possible. There is big literature on numerical band structure calculation of SLs. The most popular is the kp method based on $8\times8$ Hamiltonian.\cite{Altarelli83,Smith86,Wu85,Mailhiot86,Zakharova01,Klipstein10,Qiao12} Numerical methods make it possible to obtain results for this and even more complicated models without any approximation and make unnecessary any kind of perturbation theory. Carrier spectrum in the whole energy range can be obtained without the effective mass approximation.  The advatage of analytic results is that they present the whole picture of the phenomenon at hand in a clear form and make it easy to analyse effects of different factors.\cite{Gerchikov92,Laikhtman97,Rodina02} Contrary to numerical approach, it is not necessary to repeat all calculations to apply analytical results to another structure but it is enough to use finite formulas. Also the accuracy of numerical results in application to real structures should not be overestimated. Limitation of the accuracy comes from at least two sources: (1) Limited accuracy of experimentally measured parameters of the model\cite{Adachi,Vurgaftman01} and (2) Technological inaccuracy of heterostructures such as roughness and interdiffusion\cite{Tang87,Lee87,Egger97,Wood15} at interfaces and spacial fluctuations of composition of alloys.

The paper is organized in the following way. In the next Section we describe experiment and show that its results are incompatible with the existing theory. Because the theory is well established we feel it necessary to analyze it before suggesting any modification. This is made in Sec.\ref{sec:at} where we also point out its weak points in application to SLs. In Sec.\ref{sec:kps} we develop a modified approach and in Sec.\ref{sec:issk} we apply this approach for calculation of the in-plane SL spectrum. In Conclusion we discuss the results.

\section{Experiment}
\label{sec:exp}

Metamorphic technique of growing InAsSb structures\cite{Belenky11} was recently used for fabrication of InAsSb$_{x}$/InAsSb$_{y}$ short period SLs with band gaps varying from hundreds to 0 meV.\cite{Belenky15} Cyclotron resonance (CR) measurements were carried out in a SL with the effective band gap of $\sim10$meV with magnetic field in the growth direction (Faraday geometry).\cite{Suchalkin} Typical plot of the CR resonance energy vs square root of magnetic field $B$ is presented in Fig.1. The energies of the electron CR transition between 0th and 1st Landau levels indicated by the hollow circles are on a straight line passing through the origin. The solid triangles correspond to the transitions between electron and hole Landau levels. The y-intercept of this line gives the effective SL bandgap. The solid squares were interpreted as the electron transitions between 1st and 2nd Landau levels.\cite{Ludwig14} This transition vanishes at higher magnetic field due to depopulation of the 2nd Landau level. The linear dependence of the CR resonance energy on $\sqrt{B}$ is a clear indication on a linear character of the electron dispersion.\cite{McClure56,Haldane88,Jiang07} One can see that the energy range of the electron dispersion linearity starts at $\sim$10 meV.
\begin{figure}
\includegraphics[scale=0.6]{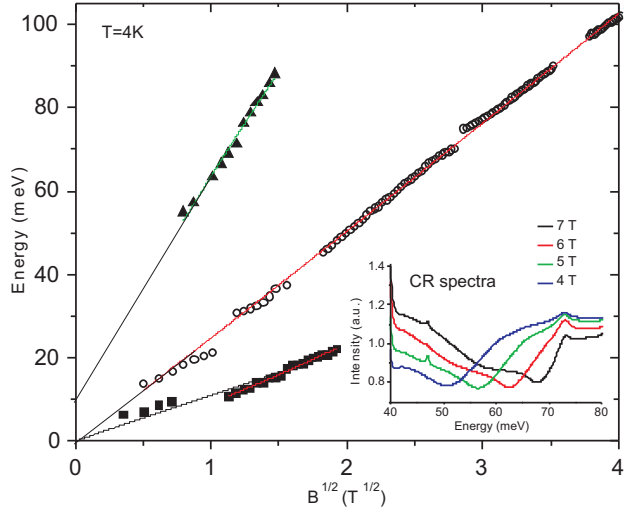}
\caption{\label{fig:cr}Absorption peaks energies vs magnetic field for 1$\mu$m thick InAsSb$_{0.3}$ (4nm)/InAsSb$_{0.75}$(2nm) SL. The magnetic field is parallel to the growth direction. Insert: electron CR peaks at different magnetic fields.}
\end{figure}

The only analytical theory of electron spectrum in SLs is based on the Kroning - Penney model. (An analytical theory based on $\bm{kp}$ Hamiltonian is developed only for narrow band constituent materials.\cite{Gerchikov92}) This model leads to the following equation
\begin{equation}
\cos qd = \cos k_{1}d_{1}\cos k_{2}d_{2} -
    \frac{1}{2} \left(\frac{m_{2}k_{1}}{m_{1}k_{2}} + \frac{m_{1}k_{2}}{m_{2}k_{1}}\right)
    \sin k_{1}d_{1}\sin k_{2}d_{2} \ ,
\label{eq:exp.1}
\end{equation}
where $d_{1}$ and $d_{2}$ are the well and barrier width, $d_{1}+d_{2}=d$, $m_{1}$ and $m_{2}$ are effective masses in the well and barrier material, $k_{1}$ and $k_{2}$ are wave vectors in the growth direction in the well and barrier defined as
\begin{equation}
k_{1} = \sqrt{\frac{2m_{1}E}{\hbar^{2}} - k_{\parallel}^{2}} \ , \hspace{1cm}
    k_{2} = \sqrt{\frac{2m_{2}(E - U)}{\hbar^{2}} - k_{\parallel}^{2}} \ ,
\label{eq:exp.2}
\end{equation}
$U$ is the height of the barrier and $\bm{k}_{\parallel}$ is the in-plane wave vector. Eqs.(\ref{eq:exp.1}) and (\ref{eq:exp.2}) define the dependence of energy $E$ on $\bm{k}_{\parallel}$ and quasi wave vector in the growth direction $q$. If $m_{1}=m_{2}$ then $E-\hbar^{2}k_{\parallel}^{2}/2m=f(q)$, i.e., the in-plane spectrum is parabolic. Difference of the masses in different layers leads to non-parabolicity of the spectrum. Electron masses in InAs and InSb differ by about two times ($m_{\rm InSb}=0.135$, $m_{\rm InAs}=0.26$\cite{Adachi,Vurgaftman01}) and this difference cannot lead to linear in-plane spectrum starting from 10meV.

Another factor that accounts for non-parbolicity of the in-plane spectrum is non-parabolicity of the bulk spectrum in the material of separate layers. It can be described by energy dependence of $m_{1}$ and $m_{2}$. This dependence becomes significant at energies of the order of the band gap in the material of any of SL layers.\cite{Kane57}  In InAsSb alloys the band gap is larger than 100 meV in the whole range of Sb concentration.\cite{Adachi,Vurgaftman01}

The one order of magnitude discrepancy between experimental and theoretical region of parabolic spectrum cannot be related to small inaccuracy of the measurement, Eq.(\ref{eq:exp.1}) or material constants. It clearly shows that something is wrong with the theory.

\section{Analysis of the theory}
\label{sec:at}

Spectrum of SLs is usually calculated with help of Kronig - Penney model that is periodic system of uniform layers of different material separated with sharp interfaces. Each layer is described with a Hamiltonian in the effective mass approximation and at interfaces carrier wave function fulfills some boundary conditions which meet symmetry limitation and current conservation. Parameters of the Hamiltonian, such as band edge energies, effective masses or Luttinger parameters for valence band are taken equal to their values in corresponding bulk material of the layer. The model gives a simple and comprehensive picture of the spectrum and is widely used.\cite{Mukherji75,Schulman81,Bastard81,Smith86,Cho87,Smith90,Ivchenko,Goncharuk05,Qiao12}

A cornerstone of this approach is the assumption that $\bm{kp}$ method, that is a basis
for the effective mass approximation, can be applied to each layer separately leading to layers' bulk spectrum and SL spectrum can be found based on these results. Some discrepancies between calculated energy levels and their measured values\cite{Mukherji75,Schulman81,Mailhiot86} can be related to uncertainties in parameters used in the calculation and these uncertainties are really considerable.\cite{Adachi,Vurgaftman01}

Primary interest in SL carrier spectrum is typically the spectrum along the growth direction, although the in-plane spectrum is also important for vertical transport because it controls the density of states. Due to separation of growth direction and in-plane variables in each layer the in-plane momentum enters SL dispersion relation as a parameter and the in-plane spectrum can be obtained in a very simple way.\cite{Bastard81,Ivchenko}

The essence of $\bm{kp}$ method is that carrier wave functions and spectrum near the band edge are calculated with help of perturbation of the state at the edge. In this paper we assume that the edges of relevant bands are at the center of the Brillouin zone that is true in III-V materials. The starting point of the method is the Hamiltonian in the basis of Bloch wave functions at the center of the zone. At the center where the carrier wave vector $\bm{k}=0$ the Hamiltonian is a diagonal matrix with diagonal elements equal energies at band edges. Away from the center of the zone kinetic energy of free electron is added to the diagonal elements. Also off-diagonal matrix elements describing coupling between bands are non-zero and proportional to wave vector $\bm{k}$ components and matrix elements of momentum $\bm{p}$ components. The Hamiltonian contains also spin-orbit interaction that is important in III-V crystals.

In the most advanced version of the method, Kane model, the coupling between the conduction and valence band is taken into account exactly while coupling with other bands is considered as a perturbation.\cite{Kane57} If this perturbation is neglected the Hamiltonian can be diagonalized exactly. In cubic crystal the spectrum is characterized by only two parameters, matrix element of a momentum component between conduction and valence band (due to cubic symmetry matrix elements of all components are equal) and spin-orbit energy splitting. This Hamiltonian is spherically symmetric and its eigenfunctions are transformed under rotations according to rotation group representations with $J=1/2$ and $J=3/2$. Perturbation by coupling with other bands leads to cubic anisotropy of the valence band that makes the Hamiltonian of the valence band identical to Kohn - Luttinger Hamiltonian.

In SLs only in-plane wave vector $\bm{k}_{\parallel}$ is a good quantum number. In the growth direction, unlike bulk, wave functions are not plane waves and new quantum numbers are the number of a subband $l$ and quasi wave vector $q$ that varies within interval $(-\pi/d,\pi/d)$ where $d$ is the SL period. If $\bm{k}_{\parallel}=0$ and the width of SL layers is much larger than the lattice constant $\bm{kp}$ Hamiltonian can be diagonalized in each layer and SL spectrum and wave functions can be obtained with help of appropriate conditions at interfaces between the layers.

If $\bm{k}_{\parallel}\neq0$ there is an additional part of the $\bm{kp}$ Hamiltonian, ${\cal H}_{\parallel}$, proportional to $\bm{k}_{\parallel}$, Eq.(\ref{eq:kps.7a}). In bulk $\bm{k}_{\parallel}\neq0$ is equivalent to rotation of the wave vector. If only conduction - valence band coupling is included the rotation does not change the spectrum and new wave functions are obtained with rotational transformation. This procedure does not work in SLs where the symmetry is uniaxial even within separate layers. At small $\bm{k}_{\parallel}$ it is possible to consider ${\cal H}_{\parallel}$ as a perturbation. Corrections to the spectrum contain (i) matrix elements of ${\cal H}_{\parallel}$ between states with $\bm{k}_{\parallel}=0$ and different $l$ and (ii) energy difference between these states. These quantities differ from corresponding quantities in bulk because of difference between wave functions and spectrum in the growth direction. Therefore the assumption that the in-plane spectrum in separate layers is the same as in bulk is baseless.

In the next section we consider a new approach to calculation of in-plane spectrum in SLs.

\section{$\bm{kp}$ equation for superlattice}
\label{sec:kps}

We study SL spectrum in Kane - type Hamiltonian and for simplicity neglect spin-orbit split band. Then in the basis
\begin{equation}
\begin{pmatrix} u_{+} \\ u_{-} \end{pmatrix} , \hspace{1cm}
    u_{+} = \begin{pmatrix} |S,1/2\rangle \\ |3/2,3/2\rangle \\ |3/2,1/2\rangle \end{pmatrix} ,
    \hspace{1cm}
    u_{-} = \begin{pmatrix} |S,-1/2\rangle \\ |3/2,-3/2\rangle \\ |3/2,-1/2\rangle \end{pmatrix} ,
\label{eq:kps.1}
\end{equation}
the Hamiltonian in each layer is
\begin{equation}
{\cal H} = \frac{\hbar^{2}k^{2}}{2m_{0}} +
\begin{pmatrix}
 E_{c} & \frac{\hbar P}{\sqrt{2}m_{0}} \ k_{+} & - \sqrt{\frac{2}{3}} \ \frac{\hbar P}{m_{0}} \ k_{z} & 0 & 0
        & - \frac{\hbar P}{\sqrt{6}m_{0}} \ k_{-} \\
 \frac{\hbar P}{\sqrt{2}m_{0}} \ k_{-} & E_{v} & 0 & 0 & 0 & 0 \\
 - \sqrt{\frac{2}{3}} \ \frac{\hbar P}{m_{0}} \ k_{z} & 0 & E_{v} & \frac{\hbar P}{\sqrt{6}m_{0}} \ k_{-} & 0 & 0 \\
 0 & 0 & \frac{\hbar P}{\sqrt{6}m_{0}} \ k_{+} & E_{c} & - \frac{\hbar P}{\sqrt{2}m_{0}} \ k_{-}
    & - \sqrt{\frac{2}{3}} \ \frac{\hbar P}{m_{0}} \ k_{z} \\
 0 & 0 & 0 & - \frac{\hbar P}{\sqrt{2}m_{0}} \ k_{+} & E_{v} & 0 \\
 - \frac{\hbar P}{\sqrt{6}m_{0}} \ k_{+} & 0 & 0 & - \sqrt{\frac{2}{3}} \ \frac{\hbar P}{m_{0}} \ k_{z} & 0 & E_{v} \\
\end{pmatrix}
\label{eq:kps.2}
\end{equation}
where $z$ is the growth direction, $k_{\pm}=k_{x}\pm ik_{y}$, $m_{0}$ is the free electron mass, $E_{c}$ and $E_{v}$ are energies of the conduction and valence band edge, $P=\langle S|p_{x}|X\rangle=\langle S|p_{y}|Y\rangle=\langle S|p_{z}|Z\rangle$ is the momentum matrix element between the conduction and valence band. In SL with layer width $d_{1}$ and $d_{2}$, $d_{1}+d_{2}=d$, the values of $E_{c}$, $E_{v}$ and $P$ are different in different layers,
\begin{subequations}
\begin{eqnarray}
&& E_{c} = E_{c1} \ , \hspace{0.5cm} E_{v} = E_{v1} \ , \hspace{0.5cm} P = P_{1}
    \hspace{1cm} nd < z < nd + d_{1} \ ,
\label{eq:kps.3a} \\
&& E_{c} = E_{c2} \ , \hspace{0.5cm} E_{v} = E_{v2} \ , \hspace{0.5cm} P = P_{2}
    \hspace{1cm} nd - d_{2} < z < nd
\label{eq:kps.3b}
\end{eqnarray}
\label{eq:kps.3}
\end{subequations}
where $n$ is the number of a period.

In Schr\"odinger equation variables are separated,
\begin{equation}
\Psi(r_{\parallel},z) = \frac{1}{\sqrt{S}} \ e^{i\bm{k}_{\parallel}\bm{r}_{\parallel}} \ \Xi(z) \ .
\label{eq:kps.4}
\end{equation}
and $z$-dependent part of the wave function satisfies the equation
\begin{equation}
{\cal H}_{\perp}\Xi + {\cal H}_{\parallel}\Xi = E\Xi \ .
\label{eq:kps.5}
\end{equation}
where
\begin{subequations}
\begin{eqnarray}
&& {\cal H}_{\perp} = \begin{pmatrix} H_{+\perp} & 0 \\ 0 &  H_{-\perp} \\\end{pmatrix} , \hspace{1cm}
    H_{-\perp} = UH_{+\perp}U \ ,
\label{eq:kps.6a} \\
&& {\cal H}_{+\perp} =
    \begin{pmatrix} E_{c} & 0 & 0 \\ 0 & E_{v} & 0 \\ 0 & 0 & E_{v} \end{pmatrix} +
    i \sqrt{\frac{2}{3}}  \ \frac{\hbar P}{m_{0}} \ R \ \frac{d}{dz} -
    \frac{\hbar^{2}}{2m_{0}} \ \frac{d^{2}}{dz^{2}} \ ,
\label{eq:kps.6b}
\end{eqnarray}
\label{eq:kps.6}
\end{subequations}
\begin{subequations}
\begin{eqnarray}
&& {\cal H}_{\parallel} =
    \begin{pmatrix} H_{+\parallel} & Q \\ Q^{\dag} &  H_{-\parallel} \\\end{pmatrix} , \hspace{1cm}
    H_{-\parallel} = UH_{+\parallel}^{*}U \ ,
\label{eq:kps.7a} \\
&& H_{+\parallel} =
    \begin{pmatrix}
 0 & \displaystyle \frac{\hbar P}{\sqrt{2}m_{0}} \ k_{+} & 0 \\
 \displaystyle \frac{\hbar P}{\sqrt{2}m_{0}} \ k_{-} & 0 & 0 \\
 \displaystyle 0 & 0 & 0 \\
    \end{pmatrix} +
    \frac{\hbar^{2}k_{\parallel}^{2}}{2m_{0}} \ , \hspace{1cm}
    Q =  \frac{\hbar P}{\sqrt{6}m_{0}} \ k_{+}
    \begin{pmatrix} 0 & 0 & 1 \\ 0 & 0 & 0 \\ - 1 & 0 & 0 \end{pmatrix} ,
\label{eq:kps.7b}
\end{eqnarray}
\label{eq:kps.7}
\end{subequations}
\begin{equation}
U = U^{\dag} = U^{-1} =
    \begin{pmatrix} 1 & 0 & 0 \\ 0 & - 1 & 0 \\ 0 & 0 & 1 \end{pmatrix} , \hspace{1cm}
    R = \begin{pmatrix} 0 & 0 & 1 \\ 0 & 0 & 0 \\ 1 & 0 & 0 \end{pmatrix} , \hspace{1cm} RU = UR \ .
\label{eq:kps.8}
\end{equation}

Eq.(\ref{eq:kps.5}) is provided with boundary conditions at interfaces that are limited by symmetry of the wave functions and conservation of current component normal to interfaces. Except boundary conditions there is also Bloch condition
\begin{equation}
\Xi(z + d) = \Xi(z)e^{iqd} \ .
\label{eq:kps.9}
\end{equation}

To study solutions to Eq.(\ref{eq:kps.5}) it is convenient to start with the case of $k_{\parallel}=0$ that is reduced to search of eigenfunctions and eigenvalues of ${\cal H}_{\perp}$. Hamiltonian ${\cal H}_{\perp}$ describes three types of carriers and each of them can be in the ground and excited states. Eigenvalues of ${\cal H}_{\perp}$ are $\eps_{\alpha,l}(q)$ where $\alpha=e$ for electrons, $\alpha=hh$ for heavy holes and $\alpha=lh$ for light holes. $l$ is the number of the subband. Near the edges of subbands the spectrum is parabolic, $\eps_{\alpha,l}(q)=\eps_{\alpha,l}(0)+\hbar^{2}(q-q_{0})^{2}/2m_{\alpha\perp,l}$ where $q_{0}=0,\pm\pi/d$ corresponds to the subband edge and $m_{0}/m_{\alpha\perp,l}\sim(P\hbar/m_{0}d)/(\eps_{\alpha,l}-\eps_{\alpha',l})$. All energy levels are double degenerate with respect to spin direction and in correspondence to the block structure of ${\cal H}_{\perp}$ eigenfunctions belonging to $\eps_{\alpha,l}(q)$ are
\begin{equation}
\Xi_{+\alpha}(z) =
    \begin{pmatrix} \Xi_{\alpha}(z) \\ 0 \end{pmatrix} , \hspace{1cm}
    \Xi_{-\alpha}(z) = \begin{pmatrix} 0 \\ U\Xi_{\alpha}(z)\end{pmatrix} ,
\label{eq:kps.10}
\end{equation}
where
\begin{equation}
H_{+\perp}\Xi_{\alpha}(z) = \eps_{\alpha,l}(q)\Xi_{\alpha}(z) \ .
\label{eq:kps.11}
\end{equation}
SL quantization brings in new energy scales: energy separation between states with the same $q$ but different subband number, the width of subbands and the width of gaps. Except extreme cases all of them are of the order of $\pi^{2}\hbar^{2}/2m_{\perp}d^{2}$.

Spectrum resulted from Eq.(\ref{eq:kps.5}) is formed under the affect of two different factors. One is the in-plane motion described by ${\cal H}_{\parallel}$ and the other is superlattice structure described by alternating layers with the boundary condition and Bloch condition. Which one of them is dominant depends on their relative contribution to the spectrum. If the energy scale introduced by ${\cal H}_{\parallel}$, $\sim(P\hbar k_{\parallel}/m_{0})^{2}/(E_{c}-E_{v})$, is larger than $\pi^{2}\hbar^{2}/2m_{\perp}d^{2}$ then it is possible to consider the superlattice structure as a perturbation and in the leading order to find the spectrum in each layer separately. This is the regular approach. A rough condition of its validity is $k_{\parallel}d\gg\pi$. Here we consider the opposite case when
\begin{equation}
k_{\parallel}d \ll \pi
\label{eq:kps.12}
\end{equation}
and ${\cal H}_{\parallel}$ in Eq.(\ref{eq:kps.5}) can be considered as a perturbation. In-plane spectrum under condition (\ref{eq:kps.12}) is calculated in the next section.

\section{In-plane spectrum at small $k_{\parallel}$}
\label{sec:issk}

Eigenfunctions of ${\cal H}_{+\perp}$, $\Xi_{s\alpha,ql}(z)$, $s=\pm$, are orthogonal and normalized,
\begin{equation}
\int_{0}^{d} \Xi_{s\alpha,ql}^{\dag}(z)\Xi_{s'\alpha',ql'}(z) dz =
    \delta_{ss'}\delta_{\alpha\alpha'}\delta_{ll'} \ ,
\label{eq:issk.1}
\end{equation}
and solution to Eq.(\ref{eq:kps.5}) can be expanded in $\Xi_{s\alpha,ql}(z)$:
\begin{equation}
\Xi(z) = \sum_{s\alpha,l} C_{s\alpha,l}\Xi_{s\alpha,ql}(z) \ .
\label{eq:issk.2}
\end{equation}
Expansion coefficients satisfy the following system of equations
\begin{equation}
(E - \eps_{\alpha,l}(q))C_{s\alpha,l} = \sum_{s'\alpha',l'}
    \langle{\Xi_{s\alpha,ql}|\cal H}_{\parallel}|\Xi_{s'\alpha',ql'}\rangle C_{s'\alpha',l'} \ .
\label{eq:issk.3}
\end{equation}
Under condition (\ref{eq:kps.12}) correction to the spectrum from matrix elements $\langle{\Xi_{s\alpha,ql}|\cal H}_{\parallel}|\Xi_{s'\alpha',ql'}\rangle$ off-diagonal with respect to the band number, $l\neq l'$, is small and in the leading approximation can be neglected. Then Eq.(\ref{eq:issk.3}) becomes
\begin{equation}
(E - \eps_{\alpha,l}(q))C_{s\alpha,l} = \sum_{s'\alpha'}
    \langle{\Xi_{s\alpha,ql}|\cal H}_{\parallel}|\Xi_{s'\alpha',ql}\rangle C_{s'\alpha',l}
\label{eq:issk.4}
\end{equation}
and in-plane spectrum in the $l$th subband can be found from the equation
\begin{equation}
\det\left(
    (E - \eps_{\alpha,l})\delta_{ss'}\delta_{\alpha\alpha'} -
    \langle{\Xi_{s\alpha,ql}|\cal H}_{\parallel}|\Xi_{s'\alpha',ql}\rangle
    \right)
    = 0 \ .
\label{eq:issk.5}
\end{equation}
The following calculation is made only for the ground state subband where $l=1$ and to shorten notations subscripts $l$ and $q$ will be omitted.

Calculation of matrix elements is facilitated by two factors. The first is the block structure of ${\cal H}_{\parallel}$, Eq.(\ref{eq:kps.7a}), that makes it possible to express them in matrix elements of $H_{+\parallel}$ and $Q$:
\begin{equation}
\langle{\Xi_{s\alpha}|\cal H}_{\parallel}|\Xi_{s'\alpha'}\rangle =
    \begin{pmatrix}
    \langle\Xi_{\alpha}|H_{+\parallel}|\Xi_{\alpha'}\rangle &
        \langle\Xi_{\alpha}|Q|\Xi_{\alpha'}\rangle \\
    \langle\Xi_{\alpha}|Q^{\dag}|\Xi_{\alpha'}\rangle &
        \langle\Xi_{\alpha}|H_{+\parallel}^{*}|\Xi_{\alpha'}\rangle
    \end{pmatrix} .
\label{eq:issk.6}
\end{equation}
The second factor is that $k_{z}\sim1/d$ and the parameter of $\bm{kp}$ method in Eq.(\ref{eq:kps.11}), $k_{z}P/m_{0}(E_{c}-E_{v})$, in each layer is indeed small because the width of the layers is much larger than the lattice constant. Therefore conduction and valence band mixing is weak in spite of electron and hole masses are strongly different from the free electron mass.\footnote{Small parameter of $\bm{kp}$ method, $\hbar Pk_{z}/m_{0}(E_{c}-E_{v})\ll1$, controls valence and conduction band mixing. Correction to the carrier energy resulted from it is $\sim(\hbar Pk_{z}/m_{0})^{2}/(E_{c}-E_{v})$. This corresponds to the inverse mass correction $\sim P^{2}/m_{0}^{2}(E_{c}-E_{v})$ that is typically much larger than $1/m_{0}$.} In the leading approximation the band mixing of wave functions can be neglected and SL wave functions can be calculated separately for each kind of carriers. So for each kind of carriers matrix function $\Xi_{\alpha}(z)$ has only one component:
\begin{equation}
\Xi_{e}(z) = \begin{pmatrix} \xi_{e}(z) \\ 0 \\ 0 \end{pmatrix} , \hspace{1cm}
    \Xi_{hh}(z) = \begin{pmatrix} 0 \\ \xi_{hh}(z) \\ 0 \end{pmatrix} , \hspace{1cm}
    \Xi_{e}(z) = \begin{pmatrix} 0 \\ 0 \\ \xi_{lh}(z) \end{pmatrix} .
\label{eq:issk.7}
\end{equation}
The explicit form of wave functions of electrons $\xi_{e}(z)$, heavy holes $\xi_{hh}(z)$ and light holes $\xi_{lh}(z)$ in SL is given in Appendix \ref{app:SLwf}. The spectrum can be found from Eqs.(\ref{eq:exp.1}) and (\ref{eq:exp.2}) with $k_{\parallel}=0$.

Calculation of blocks of matrix Eq.(\ref{eq:issk.6}) results in
\begin{subequations}
\begin{equation}
\langle\Xi_{\alpha}|H_{+\parallel}|\Xi_{\alpha'}\rangle =
    \begin{pmatrix}
    \displaystyle \frac{\hbar^{2}k_{\parallel}^{2}}{2m_{0}}
        & \displaystyle \frac{\hbar}{\sqrt{2}m_{0}} \ k_{+}P_{e,hh} & 0 \\
    \displaystyle \frac{\hbar}{\sqrt{2}m_{0}} \ k_{-} P_{hh,e}
        & \displaystyle \frac{\hbar^{2}k_{\parallel}^{2}}{2m_{0}} & 0 \\
    0 & 0 & \displaystyle \frac{\hbar^{2}k_{\parallel}^{2}}{2m_{0}} \\
    \end{pmatrix} ,
\label{eq:issk.8a}
\end{equation}
\begin{equation}
\langle\Xi_{\alpha}|Q|\Xi_{\alpha'}\rangle = \frac{\hbar k_{+}}{\sqrt{6}m_{0}}
    \begin{pmatrix}
    0          & 0 & P_{e,lh} \\
    0          & 0 & 0        \\
    - P_{lh,e} & 0 & 0        \\
    \end{pmatrix} ,
    \hspace{0.5cm}
    \langle\Xi_{\alpha}|Q^{\dag}|\Xi_{\alpha'}\rangle = \frac{\hbar k_{-}}{\sqrt{6}m_{0}}
    \begin{pmatrix}
    0        & 0 & - P_{e,lh} \\
    0        & 0 & 0          \\
    P_{lh,e} & 0 & 0          \\
    \end{pmatrix} ,
\label{eq:issk.8b}
\end{equation}
\label{eq:issk.8}
\end{subequations}
where
\begin{subequations}
\begin{eqnarray}
&& P_{e,hh} = \int_{0}^{d} \xi_{e}^{*}(z)P\xi_{hh}(z)dz =
    P_{1} \int_{0}^{d_{1}} \xi_{e}^{*}(z)\xi_{hh}(z)dz +
    P_{2} \int_{d_{1}}^{d_{2}} \xi_{e}^{*}(z)\xi_{hh}(z)dz \ ,
\label{eq:issk.9a} \\
&& P_{e,lh} = \int_{0}^{d} \xi_{e}^{*}(z)P\xi_{lh}(z)dz =
    P_{1} \int_{0}^{d_{1}} \xi_{e}^{*}(z)\xi_{lh}(z)dz +
    P_{2} \int_{d_{1}}^{d_{2}} \xi_{e}^{*}(z)\xi_{lh}(z)dz \ ,
\label{eq:issk.9b} \\
&& P_{hh,e} = P_{e,hh}^{*} \ , \hspace{1cm} P_{lh,e} = P_{e,lh}^{*} \ .
\label{eq:issk.9c}
\end{eqnarray}
\label{eq:issk.9}
\end{subequations}
$6\times6$ determinant in Eq.(\ref{eq:issk.5}) becomes block diagonal after cycle transposition rows and columns (3,4,6). Determinants of the two $3\times3$ diagonal blocks are equal so that
\begin{eqnarray}
&& \det \left(
    (E - \eps_{\alpha})\delta_{ss'}\delta_{\alpha\alpha'} -
    \langle{\Xi_{s\alpha}|\cal H}_{\parallel}|\Xi_{s'\alpha'}\rangle
        \right)
\nonumber \\ && \hspace{1cm} =
        \Bigg[
    \left(E - \eps_{e} - \frac{\hbar^{2}k_{\parallel}^{2}}{2m_{0}}\right)
    \left(E - \eps_{hh} - \frac{\hbar^{2}k_{\parallel}^{2}}{2m_{0}}\right)
    \left(E - \eps_{lh} - \frac{\hbar^{2}k_{\parallel}^{2}}{2m_{0}}\right)
\nonumber \\ && \hspace{1.5cm} -
    \frac{\hbar^{2}k_{\parallel}^{2}|P_{e,lh}|^{2}}{6m_{0}^{2}}
    \left(E - \eps_{hh} - \frac{\hbar^{2}k_{\parallel}^{2}}{2m_{0}}\right) -
    \frac{\hbar^{2}k_{\parallel}^{2}|P_{e,hh}|^{2}}{2m_{0}^{2}}
    \left(E - \eps_{lh} - \frac{\hbar^{2}k_{\parallel}^{2}}{2m_{0}}\right)
        \Bigg]^{2} \ .
\label{eq:issk.10}
\end{eqnarray}
That is the spectrum is double degenerate. The free electron energy $\hbar^{2}k_{\parallel}^{2}/2m_{0}$ at practical values of $k_{\parallel}$ is so small that it can be neglected in Eq.(\ref{eq:issk.10}).

To study the in-plane spectrum it is convenient to write the dispersion relation in the form
\begin{equation}
(E - \eps_{e})(E - \eps_{hh})(E - \eps_{lh}) =
    \frac{\hbar^{2}k_{\parallel}^{2}(|P_{e,lh}|^{2} + 3|P_{e,hh}|^{2})}{6m_{0}^{2}} \
    (E - \eps_{l-h}) \ ,
\label{eq:issk.11}
\end{equation}
where
\begin{equation}
\eps_{l-h} =
    \frac{|P_{e,lh}|^{2}\eps_{hh} + 3|P_{e,hh}|^{2}\eps_{lh}}{|P_{e,lh}|^{2} + 3|P_{e,hh}|^{2}} \ .
\label{eq:issk.12}
\end{equation}
There are following relations between these energies
\begin{equation}
\eps_{lh} < \eps_{l-h} < \eps_{hh} < \eps_{e} \ .
\label{eq:issk.13}
\end{equation}
Plot of the left hand side (cubic parabola) and right hand side (straight line) of Eq.(\ref{eq:issk.11}) as a function of $E$ is shown in Fig.2. Crossing points
of the curve and straight line correspond to solutions to Eq.(\ref{eq:issk.11}). The right crossing point corresponds to electrons, the middle crossing point corresponds to heavy holes and the left one corresponds to light holes.
\begin{figure}
\includegraphics[scale=0.6]{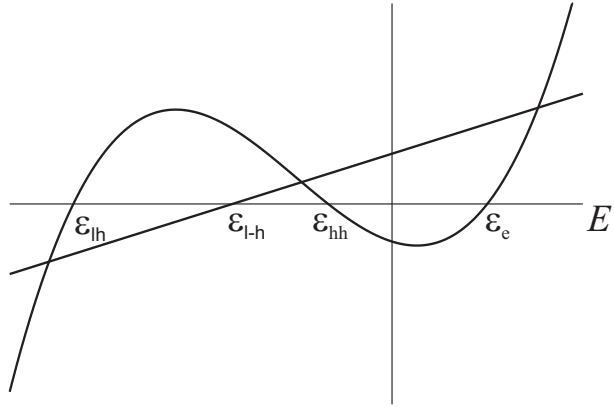}
\caption{\label{fig:sp-eq}The curve is the plot of the left side of Eq.(\ref{eq:issk.11}) and the straight line is the plot of the right hand side. Their crossing points correspond to solutions to the equation.}
\end{figure}

The right hand side of Eq.(\ref{eq:issk.11}) contains a large parameter: for the most of III-V compounds $|P_{e,hh}|^{2}/m_{0}\sim |P_{e,lh}|^{2}/m_{0}\sim P_{1}^{2}/m_{0}\approx P_{2}^{2}/m_{0}\approx10$eV. \cite{Vurgaftman01} That is with growth of $k_{\parallel}$ the prefactor in the right hand side can be quite large when $\hbar^{2}k_{\parallel}^{2}/2m_{0}$ is still small. This prefactor controls the slope of the straight line in Fig.2. With growth of $k_{\parallel}$ the straight line rotates counterclockwise. The rotation leads to growth of the electron energy and absolute value of the light hole energy without any limitation while the absolute value of the heavy hole energy grows and approaches $\eps_{l-h}$.

If $\eps_{e}-\eps_{hh}\sim\eps_{hh}-\eps_{lh}$ then asymptotic solution of Eq.(\ref{eq:issk.11}) at small $k_{\parallel}$,
\begin{equation}
\frac{\hbar^{2}k_{\parallel}^{2}}{2m_{0}} \ \frac{P^{2}}{m_{0}} \ll (\eps_{e} - \eps_{lh})^{2} \ ,
\label{eq:issk.14}
\end{equation}
gives parabolic spectrum:
\begin{subequations}
\begin{equation}
E = \eps_{e}(q) + \frac{\hbar^{2}k_{\parallel}^{2}}{2m_{e\parallel}} \ , \hspace{1cm}
    \frac{1}{m_{e\parallel}} = \frac{1}{3m_{0}^{2}} \
        \left(
    \frac{|P_{lh,e}|^{2}}{\eps_{e} - \eps_{lh}} + \frac{3|P_{e,hh}|^{2}}{\eps_{e} - \eps_{hh}}
        \right)
\label{eq:issk.15a}
\end{equation}
for electrons,
\begin{equation}
E = \eps_{hh}(q) - \frac{\hbar^{2}k_{\parallel}^{2}}{2m_{hh\parallel}} \ , \hspace{1cm}
    \frac{1}{m_{hh\parallel}} = \frac{1}{m_{0}^{2}} \ \frac{|P_{e,hh}|^{2}}{\eps_{e} - \eps_{hh}}
\label{eq:issk.15b}
\end{equation}
for heavy holes and
\begin{equation}
E = \eps_{lh}(q) - \frac{\hbar^{2}k_{\parallel}^{2}}{2m_{lh\parallel}} \ , \hspace{1cm}
    \frac{1}{m_{lh\parallel}} = \frac{1}{3m_{0}^{2}} \ \frac{|P_{e,lh}|^{2}}{\eps_{e} - \eps_{lh}}
\label{eq:issk.15c}
\end{equation}
\label{eq:issk.15}
\end{subequations}
for light holes.

A weak nonparabolicity of the electron spectrum at not very large $k_{\parallel}$ is usually described by energy dependence of the effective mass. In Eq.(\ref{eq:issk.15a}) this is reduced to replacement of $\eps_{e}$ with $E$ in the expression for the mass. If the splitting between light and heavy holes is neglected (i.e. $\eps_{lh}$ assumed to be equal $\eps_{lh}$) and the difference between momentum matrix elements in different layers $|P_{e,hh}|$ and $P_{e,lh}|$ is also neglected then the modified expression becomes identical to Eq.(1) of Ref.\cite{Wetzel96} where contribution of remote bands and spin-orbit split band is discarded.

At large $k_{\parallel}$
\begin{equation}
\frac{\hbar^{2}k_{\parallel}^{2}}{2m_{0}} \ \frac{P^{2}}{m_{0}} \gg (\eps_{e} - \eps_{lh})^{2}
\label{eq:issk.16}
\end{equation}
the energy of heavy holes is saturated (keep in mind that admixture of remote bands is neglected)
\begin{equation}
E = \frac{|P_{e,lh}|^{2}\eps_{hh} + 3|P_{e,hh}|^{2}\eps_{lh}}{|P_{e,lh}|^{2} + 3|P_{e,hh}|^{2}}
\label{eq:issk.17}
\end{equation}
while the spectrum of electrons and light holes is linear
\begin{equation}
E = \pm \frac{\hbar k_{\parallel}}{m_{0}}
    \sqrt{\frac{(|P_{e,lh}|^{2} + 3|P_{e,hh}|^{2})}{6}} + \frac{\eps_{e}}{2} +
    \frac{|P_{e,lh}|^{2}\eps_{lh} + 3|P_{e,hh}|^{2}\eps_{hh}}
    {2(|P_{e,lh}|^{2} + 3|P_{e,hh}|^{2})} \ .
\label{eq:issk.18}
\end{equation}

In case interesting for InAsSb SLs, when gap between electrons and heavy holes is anomalously small,
\begin{equation}
\eps_{e} - \eps_{hh} \ll \eps_{hh}-\eps_{lh} \ .
\label{eq:issk.19}
\end{equation}
there exists also an intermediate asymptote when
\begin{equation}
\eps_{e} - \eps_{hh} \ll |E - \eps_{e}| \ll \eps_{hh} - \eps_{lh} \ .
\label{eq:issk.20}
\end{equation}
In this energy region the first term in the right hand side of the expression
\begin{equation}
E - \eps_{l-h} = \frac{|P_{lh,e}|^{2}}{|P_{lh,e}|^{2} + 3|P_{e,hh}|^{2}} \ (E - \eps_{hh}) +
    \frac{3|P_{hh,e}|^{2}}{|P_{lh,e}|^{2} + 3|P_{e,hh}|^{2}} \ (E - \eps_{lh})
\label{eq:issk.21}
\end{equation}
can be neglected and then Eq.(\ref{eq:issk.11}) leads to
\begin{equation}
E = \frac{\eps_{e} + \eps_{hh}}{2} \pm
    \sqrt{\frac{(\eps_{e} - \eps_{hh})^{2}}{4} + \frac{\hbar^{2}k_{\parallel}^{2}|P_{hh,e}|^{2}}{2m_{0}^{2}}}
\label{eq:issk.22}
\end{equation}
where the plus corresponds the electron spectrum and minus corresponds to heavy hole spectrum. At small $k_{\parallel}$ Eq.(\ref{eq:issk.22}) gives parabolic spectrum Eqs.(\ref{eq:issk.15a}) and (\ref{eq:issk.15b}) while at large $k_{\parallel}$ the spectrum is linear,
\begin{equation}
E = \pm \frac{\hbar k_{\parallel}|P_{hh,e}|}{\sqrt{2}m_{0}} + \frac{\eps_{e} + \eps_{hh}}{2}
\label{eq:issk.23}
\end{equation}

A sketch of the whole in-plane spectrum is shown in Fig.3.
\begin{figure}
\includegraphics[scale=0.6]{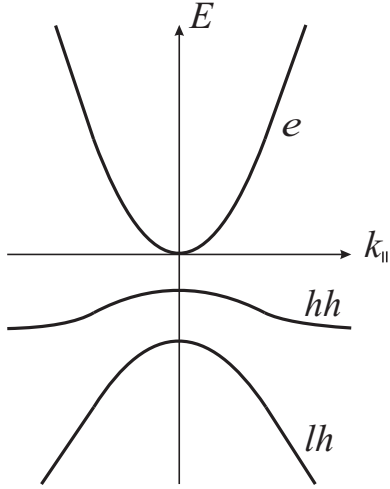}
\caption{\label{fig:sp}A sketch of the in-plane spectrum of electrons, heavy holes and light holes.}
\end{figure}

Eqs.(\ref{eq:issk.14}), (\ref{eq:issk.15}) and (\ref{eq:issk.20}) - (\ref{eq:issk.23}) show that the non-parabolic spectrum starts from the energy above electron - heavy hole gap and it is linear. These results are in agreement with the experiment.

\section{Conclusion}

Arguments presented in this paper reveal that the in-plane spectrum in SLs at not very large in-plane momentum $k_{\parallel}$ is not correctly described by the existing theory that uses effective mass approximation in each separate layer. Actually the in-plane spectrum is formed on the top of SL spectrum in the growth direction. The physical reason is that at small $k_{\parallel}$ the in-plane motion is relatively slow and before an electron makes a noticeable in-plane shift it moves across a few SL periods.

From theoretical point of view when the width of SL layers goes to infinity the SL spectrum has to go to the bulk spectrum. This indeed happens not by decrease of differences between the spectra but by shrinkage of the region of small $k_{\parallel}$ where these differences are substantial, see Eq.(\ref{eq:kps.12}).

In practical terms the limitation on $k_{\parallel}$ from above is not that strong. For period $d=10$ nm and effective mass $m\sim0.025$ the energy of the considered region has to be smaller than $\pi^{2}\hbar^{2}/2md^{2}\sim600$ meV.

The background of the in-plane spectrum is important for the spectrum character because a substantial role in its formation is played by valence - conduction band gap. The value of the gap controls the value of the effective mass, Eq.(\ref{eq:issk.15}) [similar calculation in bulk gives $1/m=(4/3)P^{2}/m_{0}^{2}(E_{c}-E_{v})$], and the size of the energy region where the spectrum is parabolic. In SL the gap can be significantly smaller than in separate layers. The reason is that the conduction and valence band gap between different layers $E_{c1}-E_{v2}$ can be much smaller than the gap in each of them, $E_{c1}-E_{v1}$ and $E_{c2}-E_{v2}$, and the SL gap is different from $E_{c1}-E_{v2}$ by quantum size effect. This is actually the case in InAsSb SLs.\cite{Belenky11} Another example is InAs/GaSb SLs close to type II. If $E_{c}-E_{v}$ is not small the results of our theory is close to the results of regular approach and may be not distinguishable in experiments.

In this paper we are not trying to explain results of the cyclotron measurements presented in Sec.\ref{sec:exp}. This is going to be done elsewhere. We can just make a rough estimate of energy difference between electron Landau levels based on the results obtained in Sec.\ref{sec:issk}. To obtain such an estimate we make use of Eq.(\ref{eq:issk.23}) and choose plus sign for electron levels. A rough value of Landau level energy can be obtained by replacing $\hbar^{2}k_{\parallel}^{2}/2$ with $(eB\hbar/c)(n+1/2)$ where $B$ is magnetic field and $n$ is the number of the level. This leads to linear dependence of the cyclotron resonance energy on $\sqrt{B}$. Making use of known value $P^{2}/m_{0}\approx10$ eV \cite{Vurgaftman01} we obtain for the transition between the first and second Landau levels $(E_{2}-E_{1})/\sqrt{B}\approx12$meV/$\sqrt{\text{T}}$ that is close to the slope 10 meV/$\sqrt{\text{T}}$ of the lowest line in Fig.1.

Finally, the results of the work can be summarized in the following points:
\begin{itemize}
\item   The size of the parabolic region in the in-plane spectrum is of the order of the conduction - valence band gap of the SL.
\item   The value of the effective mass is inverse proportional of the SL band gap.
\item   Beyond the parabolic region the electron and light hole spectrum is linear.
\item   If the gap between heavy holes and conduction band is much smaller than the gap between light holes and conduction band linear spectrum of the electron and heavy holes starts at the energy around the smaller gap.
\item   The size of the parabolic region of the spectrum and its linearity beyond this region are consistent with recent CR measurements. A rough estimate of the cyclotron resonance energy gives a value close to the experimentally measured one.
\end{itemize}

\section{Acknowledgements}

We appreciate very helpful discussions with L. D. Shvartsman. The work was supported by U.S. Army Research Office Grant W911TA-16-2-0053.

\appendix

\section{SL wave functions}
\label{app:SLwf}

We calculate SL wave functions in the frame of regular Kronig - Penney model. Let the barrier height is $U$ and effective masses in well and barrier are $m_{1\perp}$ and $m_{2\perp}$. Schr\"odinger equation
\begin{subequations}
\begin{eqnarray}
&& \frac{\hbar^{2}}{2m_{1\perp}} \ \frac{d^{2}\xi}{dz^{2}} + \eps\xi = 0 \ ,
    \hspace{1cm} nd < z < nd + d_{1} \ ,
\label{eq:SLwf.1a} \\
&& \frac{\hbar^{2}}{2m_{2\perp}} \ \frac{d^{2}\xi}{dz^{2}} + (\eps - U) \xi = 0 \ ,
    \hspace{1cm} nd - d_{2} < z < nd \ .
\label{eq:SLwf.1b}
\end{eqnarray}
\label{eq:SLwf.1}
\end{subequations}
where $d_{1}$ is the width of the well, $d_{2}$ is the width of the barrier, $d_{1}+d_{2}=d$, with boundary conditions
\begin{subequations}
\begin{eqnarray}
&& \xi(nd + 0) = \xi(nd - 0) \ , \hspace{1cm} \xi(nd + d_{1} + 0) = \xi(nd + d_{1} - 0) \ ,
\label{eq:SLwf.2a} \\
&& \frac{1}{m_{1\perp}} \left.\frac{d\xi}{dz}\right|_{nd+0} = \left.\frac{1}{m_{2\perp}} \ \frac{d\xi}{dz}\right|_{nd-0} ,
    \hspace{1cm}
    \frac{1}{m_{1\perp}} \left.\frac{d\xi}{dz}\right|_{nd+d_{1}-0} =
    \left.\frac{1}{m_{2\perp}} \ \frac{d\xi}{dz}\right|_{nd+d_{1}+0} .
\label{eq:SLwf.2b}
\end{eqnarray}
\label{eq:1dSL.gs.2}
\end{subequations}
and Bloch condition
\begin{equation}
\xi(z + d) = \xi(z)e^{iqd} \ .
\label{eq:SLwf.3}
\end{equation}
leads to well known dispersion law, Eq.(\ref{eq:exp.1}),
and wave functions
\begin{subequations}
\begin{eqnarray}
&& \xi(z) = A\cos k_{1}z + \frac{m_{1\perp}}{k_{1}} \ B\sin k_{1}z \ , \hspace{1cm} 0 < z < d_{1} \ ,
\label{eq:SLwf.4a} \\
&& \xi(z) = A\cos k_{2}z + \frac{m_{2\perp}}{k_{2}} \ B\sin k_{2}z \ , \hspace{1cm} - d_{2} < z < 0 \ .
\label{eq:SLwf.4b}
\end{eqnarray}
\label{eq:SLwf.4}
\end{subequations}
In Eqs.(\ref{eq:SLwf.3}) and (\ref{eq:SLwf.4})
\begin{equation}
k_{1} = \frac{\sqrt{2m_{1\perp}\eps}}{\hbar} \ , \hspace{1cm}
    k_{2} = \frac{\sqrt{2m_{2\perp}(\eps - U})}{\hbar} \ .
\label{eq:SLwf.5}
\end{equation}
Two constants $A$ and $B$ can be expressed in only one:
\begin{subequations}
\begin{eqnarray}
&& A =  \left(
    \frac{k_{1}}{m_{1\perp}} \ \sin k_{1}d_{1} + \frac{k_{2}}{m_{2\perp}} \ e^{iqd}\sin k_{2}d_{2}
        \right)^{-1} C \ ,
\label{eq:SLwf.6a} \\
&& B = (\cos k_{1}d_{1} - e^{iqd}\cos k_{2}d_{2})^{-1} C \ ,
\end{eqnarray}
\label{eq:SLwf.6}
\end{subequations}
and the normalization condition
\begin{equation}
\int_{0}^{d} |\xi(z)|^{2} dz = 1
\label{eq:SLwf.7}
\end{equation}
gives the following relation between the constants
\begin{eqnarray}
&& \left(\frac{d}{2} + \frac{\sin2k_{1}d_{1}}{4k_{1}} + \frac{\sin2k_{2}d_{2}}{4k_{2}}\right) |A|^{2}
\nonumber \\ && \hspace{1.2cm} +
        \left(
     \frac{m_{1\perp}^{2}d_{1}}{k_{1}^{2}} + \frac{m_{2\perp}^{2}d_{2}}{k_{2}^{2}} -
    \frac{m_{1\perp}^{2}\sin2k_{1}d_{1}}{2k_{1}^{3}} - \frac{m_{2\perp}^{2}\sin2k_{2}d_{2}}{2k_{2}^{3}}
        \right)
    \frac{|B|^{2}}{2}
\nonumber \\ && \hspace{1.2cm} +
        \left(
    \frac{1 - \cos2k_{1}d_{1}}{k_{1}^{2}} \ m_{1\perp} - \frac{1 - \cos2k_{2}d_{2}}{k_{2}^{2}} \ m_{1\perp}
        \right)
    \frac{\text{Re}(AB^{*})}{2} = 1 \ .
\label{eq:SLwf.8}
\end{eqnarray}
All relations in this Appendix are equally valid for $\eps<U$ and $\eps>U$, i.e., for real and imaginary $k_{2}$.

Wave functions of electrons, heavy holes and light holes differ by values of effective masses, the height of the barrier and the width of the wells and barriers: layers presenting wells for electrons are barriers for hole and the other way around.

\pagebreak

\begin{center}
\textbf{Figure cations}
\end{center}

Fig.1:      Absorption peaks energies vs magnetic field for 1$\mu$m thick InAsSb$_{0.3}$ (4nm)/InAsSb$_{0.75}$(2nm) SL. The magnetic field is parallel to the growth direction. Lines display the best fit to the data points. Insert: electron CR peaks at different magnetic fields.

Fig.2:      The curve is the plot of the left side of Eq.(\ref{eq:issk.11}) and the straight line is the plot of the right hand side. Their crossing points correspond to solutions to the equation.

Fig.3:      A sketch of the in-plane spectrum of electrons, heavy holes and light holes.

\end{document}